\documentclass[11pt]{article}
\pdfoutput=1

\newcommand{\INCLUDE}[1]{\includegraphics*{#1.eps}}

\usepackage{amsfonts,amsmath,amssymb,graphics,eurosym}
\usepackage{lscape}

\newcommand{\mycaption}[1]{\caption{ \small #1 } }

\newcommand{\N}{\mathbb{N}}
\newcommand{\R}{\mathbb{R}}
\newcommand{\C}{\mathbb{C}}
\newcommand{\Real}{\mathrm{Re}\,}

\newcommand{\notthis}[1]{}

\newcommand{\lgen}{\varphi}

\newcommand{\Gbar}{\overline{G}}
\newcommand{\Xbar}{\overline{X}}
\newcommand{\recov}{\Re}

\newcommand{\inv}{^{-1}}
\newcommand{\half}{\frac{1}{2}}
\newcommand{\ex}{\mathbf{E}}
\newcommand{\mykappa}{\widehat{\kappa}}
\newcommand{\kld}{\mathsf{d}_M}
\newcommand{\pderiv}[2]{\frac{\partial{#1}}{\partial{#2}}}

\newcommand{\pdderiv}[2]{\frac{\partial^2{#1}}{\partial{#2}^2}}

\begin{document}

\title{\bf Credit migration: Generating generators}
\author{Richard J. Martin\footnote{\noindent 
Department of Mathematics, Imperial College London, Exhibition Road, London SW7 2AZ, U.K. Email {\tt richard.martin1@imperial.ac.uk}} }

\maketitle

\begin{abstract}
Markovian credit migration models are a reasonably standard tool nowadays, but there are fundamental difficulties with calibrating them. We show how these are resolved using a simplified form of matrix generator and explain why risk-neutral calibration cannot be done without volatility information.
We also show how to use elementary ideas from differential geometry to make general inferences about calibration stability.

This is the longer version of a paper published in RISK (Feb 2021).
\end{abstract}

\section*{Introduction}

Credit markets face uncertain times, and the current upset caused by the coronavirus pandemic is likely to result in deterioration of the credit fundamentals for many issuers. 
Although there has been a huge rally since the end of the Global Financial Crisis, which now seems a distant memory, there have been upsets in the Eurozone (2011), and in commodity and energy markets (2014--16), while in emerging markets there have been downgrades in the sovereigns of Brazil, Russia and Turkey, and even during the benign year of 2019, the great majority of LatAm corporate names on rating-watch were potential downgrades rather than upgrades\footnote{See e.g.\ Credit Suisse Latin America Chartbook.}. 
It is worthwhile, we think, to revisit the subject of credit migration.

A standard modelling framework, introduced by Jarrow et al.~\cite{Jarrow97} over twenty years ago, uses a Markov chain, parametrised by a generator matrix, to represent the evolution of the credit rating of an issuer. A good overview of subsequent developments, including its extension to multivariate settings, is given by \cite{Bielecki11}; in terms of how it relates to the credit markets and default history, an excellent series of annual reports is produced by Standard \& Poors, of which the most recent is \cite{SandP19}. 
In the taxonomy of credit models, the ratings-based Markov chain model is best understood as a sort of reduction of the credit barrier models, in the sense that the firm value, or distance to default, is replaced by the credit rating. This reduces the dimensionality from a continuum of possible default distances to a small number of rating states, a natural idea as credit rating is a convenient way of thinking about a performing issuer.

The Markov chain model can be used in a number of ways, making it of general interest:
\begin{itemize}
\item
Econometrically, to tie potential credit losses to econometric indicators and thence to bank capital requirements \cite{Fei13};
\item
Risk management of vanilla and exotic credit portfolios;
\item
Valuation of structured credit products. It is worth bearing in mind that although the synthetic CDO market is a fraction of its former self, the CLO market is still thriving, with new CLOs being issued frequently. Further, the `reg-cap' sector (trades providing regulatory capital relief for banks) is growing, and requires the pricing and risk management of contingent credit products;
\item
Understanding the impact of rating-dependent collateral calls in XVA;
\item
Stripping the credit curve from bond prices, as a way of constructing sensibly-shaped yield curves for different ratings and tenors.  
\end{itemize}

The model can, therefore, be used in both subjective (historical) and market-implied (risk-neutral) modelling, and we shall consider both here.
Nevertheless there are several difficulties that have not been discussed in the literature, detracting from the utility of the model and limiting its application hitherto:
\begin{itemize}
\item[(I)]
The number of free parameters is uncomfortably high (with $n$ states plus default it is $n^2$);
\item[(II)]
Computation of transition probabilities requires the generator matrix to be exponentiated, but this is a trappy subject  \cite{Moler03};
\item[(III)]
There are algebraic hurdles associated with trying to make a generator matrix time-varying.
\end{itemize}
Additionally the following fundamental matters are insufficiently understood:
\begin{itemize}
\item[(IV)]
What is the best way to calibrate to an empirical (historical) transition matrix, and what level of accuracy may reasonably be asked for?
\item[(V)]
Model calibration is one of the most important disciplines in quantitative finance, and yet no serious attempt has been made to answer the following: Can we uniquely identify the risk-neutral generator from the term structure of credit spreads for each rating? If not, what extra information is needed?
\end{itemize}
We address all of these here, solving most of the above problems using a new parametrisation in which a tridiagonal generator matrix is coupled with a L\'evy stochastic time change (we call this TDST).

\notthis{
*** Limitations: deal with inverted IG curves; unduly steep curves idiosyncratically
}

\section{Tridiagonal--stochastic time change (TDST) model}
\label{sec:tdst}

We consider an $n$-states-plus-default continuous-time Markov chain and write $G\in\R^{n+1,n+1}$ for the generator matrix with the $(n+1)$th state denoting default (we will write this state as $\maltese$).
The off-diagonal elements of $G$ must be nonnegative and the rows sum to zero; the last row is all zero because default is an absorbing state. The probability of default in time $\tau$, conditional on starting at state $i$, is the $i$th element of the last column of the matrix $\exp(G\tau)$: in other words $[\exp(G\tau)]_{i\maltese}$.

Let us begin by writing the generator matrix $G\in\R^{n+1,n+1}$ in the form
\begin{equation}
G = \begin{bmatrix} H & v \\ 0 & 0 \end{bmatrix}
\label{eq:G0}
\end{equation}
where $H\in\R^{n,n}$, and $v\in\R^n$ is such that the rows of $G$ sum to zero; the offdiagonal elements of $H$, and those of $v$, must of course be nonnegative.
The $\tau$-period transition matrix is given by
\[
\exp(G\tau) =  \begin{bmatrix} \exp(H\tau) & \mathcal{E}_1(H\tau) \cdot v\tau \\ 0 & 1 \end{bmatrix}
\]
where the function $\mathcal{E}_1$ is defined by $\mathcal{E}_1(x)=(e^x-1)/x$.
As is apparent, we have used a matrix exponential, and also the function $\mathcal{E}_1$,  and it is worth making some general remarks about analytic functions of matrices.

Let $f$ be a function analytic on a domain $\mathcal{D}\subset \C$. Let $B(z_0,\rho)$ denote the open disc of centre $z_0$ and radius $\rho$, and assume that this lies inside $\mathcal{D}$. Then $f$ possesses a Taylor series $\sum_{k=0}^\infty f_k(z-z_0)^k$ convergent for $z\in B(z_0,\rho)$. It is immediate, proven for example by reduction to Jordan normal form \cite{Cohn84}, that the expansion
\[
\sum_{k=0}^\infty f_k (A-z_0 I)^k 
\]
enjoys the same convergence provided that all the eigenvalues of $A$ lie in $B(z_0,\rho)$. We can therefore write $f(A)$ without ambiguity, and will use this notation henceforth. This construction establishes the existence of the matrix exponential ($f_k=1/k!$ and $\rho=\infty$), though it does not at all make it a good computational method, because roundoff errors can be substantial\footnote{For example in computing $e^{-z}$ at $z=20$: the roundoff error is far in excess of the correct answer.}. Taylor series expansion is one of the many bad ways of calculating the matrix exponential \cite{Higham08,Moler03}---a point that is routinely ignored. A better idea is 
\begin{equation}
\exp(A) = \lim_{n\to\infty} (I+A/n)^n = \lim_{n\to\infty} (I-A/n)^{-n}
\label{eq:exp}
\end{equation}
and in each case computation is easiest when $n$ is a power of two because it can be accomplished by repeated squaring. The second expression is slightly harder to calculate as it requires a matrix inverse, but it has an advantage: for $n\in\N$,  $(I-A/n)^{-n}$ is a valid transition matrix whereas  $(I+A/n)^n$ may not be.

Having discussed the matrix exponential, we also have something to say about its inverse. It is in principle possible to write down a $\tau$-period (commonly one year) transition matrix $M$ and ask if it comes from a generator $G$, which amounts to finding the matrix logarithm of $M$. If such a generator exists then $M$ can be raised to any real positive power, and is sometimes said to be \emph{embeddable}\footnote{`Embedding': there exists a smooth map $\mathfrak{m}$ from $\R_{\ge0}$ into the space of valid transition matrices, obeying $\mathfrak{m}(s+t)=\mathfrak{m}(s)\mathfrak{m}(t)$. It is, of course, $\mathfrak{m}(t)=M^t$.}. Not all valid transition matrices are embeddable. By identifying a generator at the outset we avoid this problem.

Another important thread is the use of diagonalisation, implicit in the above discussion because we have already mentioned eigenvalues. If $A=PEP\inv$ for some invertible $P$ and diagonal $E$ then $f(A)=Pf(E)P\inv$. However, while a real matrix is generically\footnote{This means that if it isn't, it is arbitrarily close to one that is.} diagonalisable, it may be that $P$ is almost singular. There is also the minor irritation that $P,E$ may not be real. This method is another popular way of exponentiating matrices, but again one that is only reliable in certain contexts, for example symmetric matrices: more of this presently.

A simplification of the model is to assume that $H$ is tridiagonal: that is to say the only nonzero elements are those immediately above, below, or on the leading diagonal. We define a \emph{transitive} generator to be one in which all the super- and sub-diagonal elements of $H$ are strictly positive. An intransitive generator does not permit certain transitions to neighbouring states to occur: informally this causes the rating to `get stuck' and can be ruled out on fundamental grounds, though it does retain some interest as a limiting case.
Also, we define a \emph{restricted} generator to be one in which all the rows of $H$, except for the $n$th, sum to zero: equivalently, $v$ is zero except for its bottom element.
Both these simplified models have a clear connection to credit modelling as discretisations of the structural (Merton) model, in which the firm value is replaced by the credit rating as a distance-to-default measure. A restricted tridiagonal generator corresponds to a Brownian motion hitting a barrier, while an unrestricted tridiagonal generator has an additional dynamic of the firm suddenly jumping to default. 
The transitivity condition simply ensures that the volatility always exceeds zero.

It is clear that the space of restricted tridiagonal generators has $2n-1$ degrees of freedom, and the unrestricted one $3n-2$. These are substantially lower than the $n^2$ needed to define a general model.
Also, tridiagonality is useful not just for dimensionality reduction. A transitive generator can always be diagonalised, and has real eigenvalues. This is because $H=D\widetilde{H}D\inv$ with $D$ a positive diagonal matrix and $\widetilde{H}$ a \emph{symmetric} tridiagonal matrix; then $\widetilde{H}$ can be diagonalised via an orthogonal change of basis as $\widetilde{H}=QEQ'$ with $'$ denoting transpose. (The method is particularly fast for tridiagonal matrices: see e.g.~\cite[\S11.3]{NRC}.)
Thus
\[
\exp(G \tau) = 
 \begin{bmatrix} DQ\exp(E \tau)Q'D\inv & \ast \\ 0 & 1 \end{bmatrix}  .
\]
The right-hand column of the matrix, written as ``$\ast$'', need not be computed directly because it can be obtained from the rows summing to unity.
Note that the eigenvalues of $H$, i.e.\ the (diagonal) elements of $E$, are real and negative.
This makes computation of the matrix exponential straightforward\footnote{Technically, the limit of an intransitive generator will cause $D$ to become singular, and so the construction then fails, but this seems to be a theoretical issue only.}. As an aside we note the connection between tridiagonal matrices, orthogonal polynomials and Riemann-Hilbert problems \cite[Ch.2]{Deift00}.

The problem with tridiagonal generators is that they do not permit multiple downgrades over an infinitesimal time period, and so the probability of such events over short (non-infinitesimal) periods, while positive, is not high enough.
In this respect, the presence of a jump-to-default transition achieves little, because that is not the main route to default: far more likely is the occurrence of multiple downgrades, often by several notches at a time. So how do we make these sudden multiple downgrades occur?

To retain the structure above we employ a stochastic time change, replacing $\tau$ in the above equation by $\tau_t$ with $t$ denoting `real' time and $\tau_t$ `business' time. We are going to make $\tau_t$ a pure jump process, with the idea that business time occasionally takes a big leap, causing multiple transitions, and we call this model TDST (tridiagonal with stochastic time change).
Formally the process $(\tau_t)$ is to be a monotone-increasing L\'evy process\footnote{See e.g.~\cite{Schoutens03} for a general introduction.} described by the generator $\lgen$, i.e.:
\[
\ex_0[e^{u (\tau_t-\tau_0)}] \equiv e^{\lgen(u) t}.
\]
The $T$-period transition matrix is obtained by integrating over all paths $\{\tau_t\}_{t=0}^T$. As the dependence on $\tau$ is through an exponential, this is straightforward, and the effect is to replace $H$ by $\lgen(H)$ in (\ref{eq:G0}), adjusting the last column as necessary. (In context $\lgen$ will be regular in the left half-plane, so we can legitimately write the matrix function $\lgen(H)$.)
Accordingly the generator and $T$-period transition matrix are
\begin{equation}
G_\lgen = \begin{bmatrix} \lgen(H) & \ast \\ 0 & 0 \end{bmatrix}
; \qquad
M_\lgen(T) = \begin{bmatrix} \exp(\lgen(H)T) & \ast \\ 0 & 1 \end{bmatrix}
= \begin{bmatrix} DQ\cdot \exp(\lgen(E) T) \cdot Q'D\inv & \ast \\ 0 & 1 \end{bmatrix}.
\label{eq:G2}
\end{equation}
We can impose that $\tau_t/t$ have unit mean (as otherwise the model is over-specified because $H$ and $\tau$ can be scaled in opposite directions), so $\lgen'(0)=1$. 
An obvious choice is the CMY process
\begin{equation}
\lgen(u) = \frac{\beta}{\gamma} \big(1-(1-u/\beta)^\gamma\big), \qquad \beta>0, \quad \gamma<1
\label{eq:cmy}
\end{equation}
which has as special cases the Inverse Gaussian process and the Gamma process, respectively ($\gamma=\half,0$):
\[
\lgen(u) = 2 \beta (1-\sqrt{1-u/\beta}), \qquad
\lgen(u) = - \beta  \log (1-u/\beta).
\]
This retains the connection with structural (Merton) modelling, in that we have a discretisation of a L\'evy process hitting a barrier---a standard idea in credit risk modelling \cite{Lipton02c,Martin09a,Martin10b,Dalessandro11}.  
An incidental remark about the Gamma process is that it allows the matrix exponential $\lgen(H)$ to be calculated in an alternative simple way, as noted in the Appendix.

It is perhaps worth emphasising that despite our using the term `stochastic time change' this model specification still has a \emph{static} generator. The stochastic time change serves only as a mechanism for generating multiple downgrades in an infinitesimal time period using a tridiagonal generator matrix.
The number of parameters for this model is $2n-1+n_\lgen$, where $n_\lgen$ ($=2$ here) is the number needed to specify $\varphi$.

\section{Calibration to historical transition matrix}
\label{sec:hc}

Israel et al.~\cite{Israel00} spend some time showing that empirical transition matrices are not necessarily embeddable (q.v.). In fact it is only necessary to test whether the empirical matrix is \emph{statistically likely} to be so. Taking S\&P's data in \cite[Tables~21,23]{SandP19}, we ask if we can find a TDST model that fits well enough.
 
There are a number of issues to consider. The first is the right measure of closeness of fit.
From the perspective of statistical theory the most appropriate is the Kullback--Leibler divergence\footnote{A Taylor series expansion around $p=q$ gives the well-known chi-squared measure of ``$(O-E)^2/E$, summed'' ($O$ observed, $E$ expected)---this can also be used to define the fitting error, and it gives similar results.}:
\begin{equation}
\kld = \sum_{i=1}^n\sum_{j=1}^{n+1} p_{ij} \ln \frac{p_{ij}}{q_{ij}} 
\label{eq:kld}
\end{equation}
where $p_{ij}$ are the historical probabilities of transition from state $i$ to $j$ (with $n+1$ denoting default) and $q_{ij}$ the model ones. This is essentially the log likelihood ratio statistic, i.e.\ the logarithm of the ratio of the likelihood of the model compared with that of the maximum likelihood estimator \cite[Ch.4]{Cox74}.
Note that $\kld$ is not symmetrical in $p,q$: nor should it be. To see why, consider first the effect of letting $q_{ij}\to0$ with $p_{ij}>0$: this means that the event has been observed but that the model probability is zero, an error that should be heavily penalised. On the other hand, $p_{ij}\to0$ with $q_{ij}>0$ simply means that the model is attaching positive probability to an event that has not been observed, which is not a major objection.

Empirical transition data reveal a problem: the probability of transiting from rated to unrated (`NR') is quite high. A standard idea is to distribute this probability pro rata amongst the elements of each row, excepting the transition to default: doing so retains, as it must, the property that each row sum to unity. However, this is not the only way of making the adjustment, and although the `NR' problem  is unfortunate, it allows for, or cannot rule out, considerable leeway in the fitting---which is why attempts to exactly fit the matrix are misguided. Another issue is that the standard deviation of low empirical probabilities is proportionally quite high. The finiteness of the samples used to construct empirical transition matrices introduces considerable relative uncertainties.

We then minimise $\kld$ with respect to the elements of $H$ (and the bottom element of $v$) and the parameters $\beta,\gamma$ in (\ref{eq:cmy}). 
Optimisation of the 7-state model takes a fraction of a second and the results are shown in Figure~\ref{fig:sp7}. It is seen that the fit is good, and furthermore the fitting errors are small by comparison with the uncertainties introduced by the `NR' problem. The errors are also typically smaller than the standard errors given by S\&P, which are in \cite[Table~21]{SandP19} but omitted here\footnote{One has to be careful in interpreting these, for a reason that we have already touched on. Certain transitions have never been observed, and for these S\&P show the probability and standard error as zero. Now if the true frequency is zero, then the mean and standard deviation of the observed frequency will be zero. But what we really want is an estimate of the true frequency and the standard deviation \emph{of that estimate}. Clearly if Bayesian inference is followed then the latter is not zero, because the true frequency may be $>0$.}.

If we pass to an 18-state model (AAA, AA+, AA, \ldots, CCC+, CCC) we still have a large number of parameters even after the reduction to a TDST model (324 to around 36). It is possible to fit these, but some further simplification proves advantageous. Suppose that the free parameters in the matrix $H$ are the sub/superdiagonal elements for the following transitions only: $\mbox{AA}\to\mbox{AA}\pm$, $\mbox{A}\to\mbox{A}\pm$, $\mbox{BBB}\to\mbox{BBB}\pm$, $\mbox{BB}\to\mbox{BB}\pm$, $\mbox{B}\to\mbox{B}\pm$,  $\mbox{CCC}\to\mbox{CCC}+$, and also the bottom element of $v$, giving $\mbox{CCC}\to\maltese$. (Hence there are 12 in all.) The values for intermediate states ($\mbox{AA}-\to\mbox{AA}/\mbox{A}+$ and so on) are found by logarithmic interpolation\footnote{For convenience $\mbox{AA+}\to\mbox{AAA}$ is taken to be the same as $\mbox{AA}\to\mbox{AA}+$, and $\mbox{AAA}\to\mbox{AA+}$ and $\mbox{AA}+\to\mbox{AA}$ the same as $\mbox{AA}\to\mbox{AA}-$.}.
A minor complication is that the S\&P data group CCC+, CCC and lower ratings into one. We have ignored any rating lower than CCC because in practice transition through these states is very rapid, and default almost always ensues \cite[Chart~10]{SandP19}.
Again the results are good: see Figure~\ref{fig:sp18f}.

\begin{figure}
\small
\begin{tabular}{ll}
(i) & \begin{tabular}[t]{l|rrrrrrrrr}
(\%) & AAA & AA & A & BBB & BB & B & CCC & $\maltese$ & NR
 \\
 \hline
AAA & $86.99$ & $9.12$ & $0.53$ & $0.05$ & $0.08$ & $0.03$ & $0.05$ & $0.00$ & $3.15$ \\
AA & $0.50$ & $87.06$ & $7.85$ & $0.49$ & $0.05$ & $0.06$ & $0.02$ & $0.02$ & $3.94$ \\
A & $0.03$ & $1.69$ & $88.17$ & $5.16$ & $0.29$ & $0.12$ & $0.02$ & $0.06$ & $4.48$ \\
BBB & $0.01$ & $0.09$ & $3.42$ & $86.04$ & $3.62$ & $0.46$ & $0.11$ & $0.17$ & $6.10$ \\
BB & $0.01$ & $0.03$ & $0.11$ & $4.83$ & $77.50$ & $6.65$ & $0.55$ & $0.65$ & $9.67$ \\
B & $0.00$ & $0.02$ & $0.08$ & $0.17$ & $4.93$ & $74.53$ & $4.42$ & $3.44$ & $12.41$ \\
CCC & $0.00$ & $0.00$ & $0.11$ & $0.20$ & $0.59$ & $13.21$ & $43.51$ & $26.89$ & $15.50$ \\
\end{tabular}

\\
\mbox{}
\\
(ii) & \begin{tabular}[t]{l|rrrrrrrr}
(\%) & AAA & AA & A & BBB & BB & B & CCC & $\maltese$
 \\
 \hline
AAA & $89.82$ & $9.42$ & $0.55$ & $0.05$ & $0.08$ & $0.03$ & $0.05$ & $0.00$ \\
AA & $0.52$ & $90.64$ & $8.17$ & $0.51$ & $0.05$ & $0.06$ & $0.02$ & $0.02$ \\
A & $0.03$ & $1.77$ & $92.29$ & $5.40$ & $0.30$ & $0.13$ & $0.02$ & $0.06$ \\
BBB & $0.01$ & $0.10$ & $3.64$ & $91.62$ & $3.85$ & $0.49$ & $0.12$ & $0.17$ \\
BB & $0.01$ & $0.03$ & $0.12$ & $5.35$ & $85.86$ & $7.37$ & $0.61$ & $0.65$ \\
B & $0.00$ & $0.02$ & $0.09$ & $0.20$ & $5.66$ & $85.52$ & $5.07$ & $3.44$ \\
CCC & $0.00$ & $0.00$ & $0.14$ & $0.25$ & $0.75$ & $16.76$ & $55.21$ & $26.89$ \\
\end{tabular}

\\
\mbox{}
\\
(iii) & \begin{tabular}[t]{l|rrrrrrrr}
(\%) & AAA & AA & A & BBB & BB & B & CCC & $\maltese$
 \\
 \hline
AAA & $89.69$ & $8.90$ & $1.08$ & $0.23$ & $0.05$ & $0.02$ & $0.00$ & $0.03$ \\
AA & $0.56$ & $91.12$ & $7.49$ & $0.66$ & $0.09$ & $0.04$ & $0.00$ & $0.04$ \\
A & $0.02$ & $1.83$ & $92.44$ & $5.21$ & $0.33$ & $0.09$ & $0.01$ & $0.07$ \\
BBB & $0.00$ & $0.11$ & $3.64$ & $91.55$ & $4.03$ & $0.47$ & $0.04$ & $0.16$ \\
BB & $0.00$ & $0.02$ & $0.30$ & $5.21$ & $85.68$ & $7.69$ & $0.46$ & $0.63$ \\
B & $0.00$ & $0.01$ & $0.06$ & $0.43$ & $5.43$ & $85.43$ & $5.69$ & $2.96$ \\
CCC & $0.00$ & $0.00$ & $0.02$ & $0.12$ & $0.96$ & $16.72$ & $55.06$ & $27.12$ \\
\end{tabular}

\\
\mbox{}
\\
& \begin{tabular}[t]{l|rrrrrrrr}
(\%) & AAA & AA & A & BBB & BB & B & CCC & $\maltese$
 \\
 \hline
AAA & $-10.91$ & $9.84$ & $0.78$ & $0.19$ & $0.04$ & $0.02$ & $0.00$ & $0.02$ \\
AA & $0.62$ & $-9.42$ & $8.16$ & $0.49$ & $0.08$ & $0.03$ & $0.00$ & $0.04$ \\
A & $0.01$ & $2.00$ & $-8.05$ & $5.66$ & $0.24$ & $0.08$ & $0.01$ & $0.06$ \\
BBB & $0.00$ & $0.08$ & $3.95$ & $-9.08$ & $4.54$ & $0.33$ & $0.03$ & $0.14$ \\
BB & $0.00$ & $0.02$ & $0.22$ & $5.87$ & $-15.88$ & $8.99$ & $0.27$ & $0.51$ \\
B & $0.00$ & $0.01$ & $0.05$ & $0.30$ & $6.35$ & $-16.91$ & $8.28$ & $1.93$ \\
CCC & $0.00$ & $0.00$ & $0.02$ & $0.09$ & $0.56$ & $24.34$ & $-60.85$ & $35.84$ \\
\end{tabular}

\\
\mbox{}
\\
(iv) & \begin{tabular}[t]{l|rrr}
  & $\uparrow$ &  0  & $\downarrow$
 \\
 \hline
AAA & $$ & $-0.1371$ & $0.1371$ \\
AA & $0.0086$ & $-0.1184$ & $0.1098$ \\
A & $0.0269$ & $-0.1024$ & $0.0755$ \\
BBB & $0.0527$ & $-0.1172$ & $0.0646$ \\
BB & $0.0835$ & $-0.2179$ & $0.1344$ \\
B & $0.0949$ & $-0.2434$ & $0.1485$ \\
CCC & $0.4364$ & $-1.0281$ & $0.5918$ \\
\end{tabular}
 \begin{tabular}[t]{l|r}
  & value
 \\
 \hline
$\gamma$ & $0.8154$ \\
$\beta$ & $0.0241$ \\
\end{tabular}

\end{tabular}
\mycaption{
S\&P 7-state historical transition matrix: (i) raw \cite[Table~21]{SandP19}, (ii) adjusted for transition to unrated (NR); 
(iii) TDST fitted one-year transition matrix $M_\lgen(1)$, and its corresponding generator $G_\lgen$, as per eq.(\ref{eq:G2});
(iv) parameters, i.e.\ the subdiagonal (upgrade, `$\uparrow$'), leading diagonal (rating unchanged, `0'), and superdiagonal (downgrade, `$\downarrow$') in eq.(\ref{eq:G0}), and the parameters $\beta,\gamma$ specifying $\lgen$ in eq.(\ref{eq:cmy}).
}
\label{fig:sp7}
\end{figure}


\begin{landscape}
\begin{figure*}
\scriptsize
\begin{tabular}{l}
[Actual]
\\
\begin{tabular}[t]{l|rrrrrrrrrrrrrrrrrr}
(\%) & AAA & AA+ & AA & AA-- & A+ & A & A-- & BBB+ & BBB & BBB-- & BB+ & BB & BB-- & B+ & B & B-- & CCC & $\maltese$
 \\
 \hline
AAA & $89.82$ & $6.08$ & $2.62$ & $0.71$ & $0.17$ & $0.25$ & $0.13$ & $0.00$ & $0.05$ & $0.00$ & $0.03$ & $0.05$ & $0.00$ & $0.00$ & $0.03$ & $0.00$ & $0.05$ & $0.00$ \\
AA+ & $2.43$ & $80.90$ & $11.44$ & $3.73$ & $0.74$ & $0.35$ & $0.20$ & $0.05$ & $0.10$ & $0.05$ & $0.00$ & $0.00$ & $0.00$ & $0.00$ & $0.00$ & $0.00$ & $0.00$ & $0.00$ \\
AA & $0.44$ & $1.38$ & $83.91$ & $8.92$ & $2.90$ & $1.23$ & $0.40$ & $0.41$ & $0.14$ & $0.08$ & $0.05$ & $0.03$ & $0.02$ & $0.02$ & $0.00$ & $0.02$ & $0.05$ & $0.02$ \\
AA-- & $0.04$ & $0.11$ & $3.98$ & $81.97$ & $10.23$ & $2.34$ & $0.65$ & $0.27$ & $0.16$ & $0.07$ & $0.03$ & $0.00$ & $0.00$ & $0.03$ & $0.08$ & $0.00$ & $0.00$ & $0.03$ \\
A+ & $0.00$ & $0.06$ & $0.47$ & $4.70$ & $81.76$ & $9.28$ & $2.28$ & $0.65$ & $0.37$ & $0.09$ & $0.06$ & $0.09$ & $0.01$ & $0.07$ & $0.03$ & $0.00$ & $0.00$ & $0.05$ \\
A & $0.04$ & $0.04$ & $0.23$ & $0.44$ & $5.59$ & $82.25$ & $7.16$ & $2.55$ & $0.93$ & $0.29$ & $0.10$ & $0.12$ & $0.07$ & $0.09$ & $0.02$ & $0.00$ & $0.01$ & $0.06$ \\
A-- & $0.04$ & $0.01$ & $0.06$ & $0.17$ & $0.46$ & $6.80$ & $81.28$ & $7.74$ & $2.16$ & $0.62$ & $0.15$ & $0.15$ & $0.13$ & $0.12$ & $0.02$ & $0.01$ & $0.03$ & $0.06$ \\
BBB+ & $0.00$ & $0.01$ & $0.06$ & $0.06$ & $0.22$ & $0.83$ & $7.58$ & $79.56$ & $8.62$ & $1.73$ & $0.39$ & $0.32$ & $0.15$ & $0.17$ & $0.11$ & $0.02$ & $0.06$ & $0.10$ \\
BBB & $0.01$ & $0.01$ & $0.04$ & $0.03$ & $0.12$ & $0.34$ & $1.10$ & $8.36$ & $80.30$ & $6.63$ & $1.49$ & $0.64$ & $0.30$ & $0.24$ & $0.12$ & $0.04$ & $0.06$ & $0.16$ \\
BBB-- & $0.01$ & $0.01$ & $0.02$ & $0.05$ & $0.06$ & $0.16$ & $0.30$ & $1.31$ & $9.99$ & $77.55$ & $5.98$ & $2.31$ & $0.93$ & $0.40$ & $0.25$ & $0.18$ & $0.24$ & $0.24$ \\
BB+ & $0.04$ & $0.00$ & $0.00$ & $0.03$ & $0.01$ & $0.10$ & $0.10$ & $0.45$ & $1.84$ & $12.61$ & $70.86$ & $8.25$ & $2.98$ & $1.10$ & $0.62$ & $0.26$ & $0.42$ & $0.32$ \\
BB & $0.00$ & $0.00$ & $0.04$ & $0.01$ & $0.00$ & $0.07$ & $0.06$ & $0.19$ & $0.55$ & $2.28$ & $10.64$ & $71.77$ & $9.32$ & $2.49$ & $1.13$ & $0.34$ & $0.59$ & $0.53$ \\
BB-- & $0.00$ & $0.00$ & $0.00$ & $0.01$ & $0.01$ & $0.01$ & $0.06$ & $0.11$ & $0.26$ & $0.40$ & $1.98$ & $10.86$ & $70.76$ & $9.47$ & $3.44$ & $0.92$ & $0.76$ & $0.95$ \\
B+ & $0.00$ & $0.01$ & $0.00$ & $0.03$ & $0.00$ & $0.03$ & $0.07$ & $0.06$ & $0.06$ & $0.12$ & $0.36$ & $1.65$ & $9.27$ & $71.09$ & $10.41$ & $2.87$ & $1.95$ & $2.01$ \\
B & $0.00$ & $0.00$ & $0.01$ & $0.01$ & $0.00$ & $0.03$ & $0.05$ & $0.02$ & $0.07$ & $0.03$ & $0.14$ & $0.29$ & $1.32$ & $8.86$ & $71.21$ & $10.08$ & $4.46$ & $3.41$ \\
B-- & $0.00$ & $0.00$ & $0.00$ & $0.00$ & $0.02$ & $0.04$ & $0.00$ & $0.08$ & $0.06$ & $0.12$ & $0.11$ & $0.16$ & $0.56$ & $2.79$ & $12.17$ & $63.44$ & $13.70$ & $6.75$ \\
CCC & $0.00$ & $0.00$ & $0.00$ & $0.00$ & $0.04$ & $0.00$ & $0.10$ & $0.08$ & $0.10$ & $0.08$ & $0.04$ & $0.22$ & $0.49$ & $1.32$ & $3.30$ & $12.14$ & $55.21$ & $26.89$ \\
\end{tabular}

\\
\mbox { } \\ \mbox{}
[Fitted]
\\
\begin{tabular}[t]{l|rrrrrrrrrrrrrrrrrrr}
(\%) & AAA & AA+ & AA & AA-- & A+ & A & A-- & BBB+ & BBB & BBB-- & BB+ & BB & BB-- & B+ & B & B-- & CCC+ & CCC & $\maltese$
 \\
 \hline
AAA & $85.65$ & $9.49$ & $2.50$ & $1.06$ & $0.56$ & $0.31$ & $0.18$ & $0.11$ & $0.06$ & $0.03$ & $0.02$ & $0.01$ & $0.01$ & $0.01$ & $0.00$ & $0.00$ & $0.00$ & $0.00$ & $0.01$ \\
AA+ & $2.64$ & $83.70$ & $9.19$ & $2.36$ & $0.99$ & $0.49$ & $0.27$ & $0.15$ & $0.09$ & $0.04$ & $0.02$ & $0.01$ & $0.01$ & $0.01$ & $0.01$ & $0.00$ & $0.00$ & $0.00$ & $0.01$ \\
AA & $0.19$ & $2.56$ & $83.90$ & $9.12$ & $2.32$ & $0.92$ & $0.45$ & $0.24$ & $0.13$ & $0.06$ & $0.03$ & $0.02$ & $0.01$ & $0.01$ & $0.01$ & $0.00$ & $0.00$ & $0.00$ & $0.02$ \\
AA-- & $0.03$ & $0.25$ & $3.44$ & $83.33$ & $9.04$ & $2.19$ & $0.86$ & $0.41$ & $0.21$ & $0.10$ & $0.05$ & $0.03$ & $0.02$ & $0.01$ & $0.01$ & $0.01$ & $0.00$ & $0.00$ & $0.03$ \\
A+ & $0.01$ & $0.05$ & $0.44$ & $4.52$ & $82.58$ & $8.78$ & $2.09$ & $0.79$ & $0.36$ & $0.16$ & $0.08$ & $0.04$ & $0.03$ & $0.02$ & $0.01$ & $0.01$ & $0.00$ & $0.00$ & $0.03$ \\
A & $0.00$ & $0.02$ & $0.11$ & $0.73$ & $5.82$ & $81.35$ & $8.61$ & $1.99$ & $0.73$ & $0.29$ & $0.13$ & $0.07$ & $0.04$ & $0.03$ & $0.02$ & $0.01$ & $0.01$ & $0.00$ & $0.05$ \\
A-- & $0.00$ & $0.01$ & $0.04$ & $0.21$ & $1.01$ & $6.30$ & $80.83$ & $8.47$ & $1.90$ & $0.60$ & $0.25$ & $0.12$ & $0.07$ & $0.05$ & $0.04$ & $0.02$ & $0.01$ & $0.00$ & $0.07$ \\
BBB+ & $0.00$ & $0.00$ & $0.02$ & $0.08$ & $0.31$ & $1.19$ & $6.91$ & $80.28$ & $8.31$ & $1.66$ & $0.56$ & $0.24$ & $0.13$ & $0.08$ & $0.06$ & $0.04$ & $0.02$ & $0.01$ & $0.11$ \\
BBB & $0.00$ & $0.00$ & $0.01$ & $0.04$ & $0.13$ & $0.39$ & $1.41$ & $7.57$ & $79.69$ & $7.70$ & $1.65$ & $0.57$ & $0.28$ & $0.17$ & $0.11$ & $0.06$ & $0.03$ & $0.01$ & $0.17$ \\
BBB-- & $0.00$ & $0.00$ & $0.00$ & $0.02$ & $0.06$ & $0.17$ & $0.50$ & $1.70$ & $8.65$ & $76.78$ & $8.40$ & $1.85$ & $0.74$ & $0.38$ & $0.24$ & $0.13$ & $0.06$ & $0.02$ & $0.29$ \\
BB+ & $0.00$ & $0.00$ & $0.00$ & $0.01$ & $0.04$ & $0.09$ & $0.23$ & $0.63$ & $2.06$ & $9.32$ & $73.84$ & $9.08$ & $2.33$ & $0.94$ & $0.51$ & $0.26$ & $0.12$ & $0.04$ & $0.48$ \\
BB & $0.00$ & $0.00$ & $0.00$ & $0.01$ & $0.02$ & $0.05$ & $0.12$ & $0.29$ & $0.77$ & $2.25$ & $9.97$ & $70.38$ & $10.54$ & $2.76$ & $1.19$ & $0.53$ & $0.24$ & $0.08$ & $0.79$ \\
BB-- & $0.00$ & $0.00$ & $0.00$ & $0.00$ & $0.01$ & $0.03$ & $0.06$ & $0.14$ & $0.33$ & $0.76$ & $2.17$ & $8.96$ & $70.16$ & $11.11$ & $3.15$ & $1.17$ & $0.48$ & $0.16$ & $1.28$ \\
B+ & $0.00$ & $0.00$ & $0.00$ & $0.00$ & $0.01$ & $0.01$ & $0.03$ & $0.07$ & $0.15$ & $0.30$ & $0.68$ & $1.83$ & $8.67$ & $69.83$ & $11.91$ & $3.04$ & $1.07$ & $0.34$ & $2.05$ \\
B & $0.00$ & $0.00$ & $0.00$ & $0.00$ & $0.00$ & $0.01$ & $0.02$ & $0.04$ & $0.07$ & $0.14$ & $0.27$ & $0.56$ & $1.76$ & $8.52$ & $70.18$ & $11.40$ & $2.88$ & $0.79$ & $3.36$ \\
B-- & $0.00$ & $0.00$ & $0.00$ & $0.00$ & $0.00$ & $0.00$ & $0.01$ & $0.02$ & $0.04$ & $0.07$ & $0.13$ & $0.25$ & $0.65$ & $2.17$ & $11.38$ & $65.48$ & $11.44$ & $2.41$ & $5.91$ \\
CCC+ & $0.00$ & $0.00$ & $0.00$ & $0.00$ & $0.00$ & $0.00$ & $0.01$ & $0.01$ & $0.02$ & $0.04$ & $0.07$ & $0.13$ & $0.31$ & $0.89$ & $3.33$ & $13.27$ & $59.82$ & $10.22$ & $11.87$ \\
CCC & $0.00$ & $0.00$ & $0.00$ & $0.00$ & $0.00$ & $0.00$ & $0.00$ & $0.01$ & $0.01$ & $0.02$ & $0.03$ & $0.06$ & $0.14$ & $0.37$ & $1.23$ & $3.77$ & $13.75$ & $50.26$ & $30.34$ \\
\end{tabular}

\end{tabular}
\end{figure*}
\end{landscape}

\clearpage

\begin{figure}
\centering
\begin{tabular}{l}
\begin{tabular}[t]{l|rrr}
  & $\uparrow$ &  0  & $\downarrow$
 \\
 \hline
AAA & $$ & $-0.3014$ & $0.3014$ \\
AA+ & $0.0838$ & $-0.3852$ & $0.3014$ \\
AA & $0.0838$ & $-0.3852$ & $0.3014$ \\
AA-- & $0.1136$ & $-0.4216$ & $0.3080$ \\
A+ & $0.1539$ & $-0.4686$ & $0.3147$ \\
A & $0.2085$ & $-0.5301$ & $0.3215$ \\
A-- & $0.2351$ & $-0.5598$ & $0.3248$ \\
BBB+ & $0.2650$ & $-0.5930$ & $0.3280$ \\
BBB & $0.2987$ & $-0.6301$ & $0.3313$ \\
BBB-- & $0.3722$ & $-0.7898$ & $0.4176$ \\
BB+ & $0.4636$ & $-0.9901$ & $0.5264$ \\
BB & $0.5776$ & $-1.2412$ & $0.6636$ \\
BB-- & $0.5645$ & $-1.2717$ & $0.7072$ \\
B+ & $0.5516$ & $-1.3053$ & $0.7537$ \\
B & $0.5391$ & $-1.3423$ & $0.8032$ \\
B-- & $0.8021$ & $-1.8315$ & $1.0294$ \\
CCC+ & $1.1934$ & $-2.5126$ & $1.3192$ \\
CCC & $1.7755$ & $-3.4662$ & $1.6907$ \\
\end{tabular}

\begin{tabular}[t]{l|r}
  & value
 \\
 \hline
$\gamma$ & $0.5152$ \\
$\beta$ & $0.0457$ \\
\end{tabular}

\end{tabular}
\mycaption{
As above but with the 18-state model. One-year transition matrix \cite[Table~23]{SandP19}, TDST fit, and parameters.
}
\label{fig:sp18f}
\end{figure}

\clearpage

\section{Making the generator time-varying}
\label{sec:tv}

\subsection{Theoretical considerations}
\label{sec:tvtheor}

We start by pointing out a trap that is very easy to fall into. When the generator varies with time, the transition matrix is \emph{not} given by the following expression (or its expectation conditional on information known at time $0$, for a stochastically varying $G$):
\begin{equation}
M(T) \stackrel{??}{=} \exp \left( \int_0^T G_t \, dt \right) \qquad \mbox{(Wrong in general)}.
\label{eq:wrong}
\end{equation}
Instead, the so-called \emph{ordered exponential}, sometimes written `$\mathcal{T}\exp$', must be used\footnote{To see where the problem is, suppose that $G_t$ equals $G^\textrm{i}$ for $t\in[0,T/2)$ and $G^\textrm{ii}$ for $t\in[T/2,T]$, both constant matrices. Then the period-$T$ transition matrix is $\exp(G^\textrm{i} T/2) \exp(G^\textrm{ii} T/2) \ne \exp(G^\textrm{i} T/2 + G^\textrm{ii} T/2 )$ because one needs commutativity but might not have it. See Wikipedia page on `ordered exponential'.}. This can be thought of as
\[
\lim_{N\to\infty} \Big[ (I+G_{0}\, \delta t)(I+G_{\delta t}\, \delta t)(I+G_{2\delta t}\, \delta t) \cdots (I+G_{(N-1)\delta t}\, \delta t) \Big]  , \qquad \delta t = T/N.
\]
What is beguiling about (\ref{eq:wrong}) is its simplicity, its `obvious' connection to standard ideas on rates modelling, and that its RHS is a valid transition matrix, albeit the wrong one.

We can invoke the Feynman-Kac representation, which is that if $X_t$ obeys the SDE 
\[
dX_t = \mu(X_t) \, dt + \sigma(X_t) \, dW_t,
\]
and $G$ is a function of $x$ only, then the matrix $M$ obeys the backward equation
\[
\pderiv{M}{t} =  G(x)M + \mu(x) \pderiv{M}{x} + \frac{\sigma^2(x)}{2} \pdderiv{M}{x}, \qquad M(0)=I;
\]
but this equation is unlikely to have a closed-form solution, so has to be solved numerically e.g.\ on a trinomial tree.

Another route that is likely to cause difficulties is an over-reliance on diagonalisation methods. We have said that if we can write $G=PEP\inv$ then it may be easy to exponentiate $G$, depending on how well-conditioned $P$ is.
To make $G$ time-varying it is superficially attractive to allow $E$ to vary while fixing $P$. The problem with that is that there is no guarantee that a valid transition matrix is thereby produced.

There is one case which, interestingly, links to both these ideas. That is where the effect of time variation is simply to scale $G$ by a factor, i.e.\ $G_t=X_t \Gbar$ with $\Gbar$ a static matrix and $X_t$ some scalar positive process which may as well have unit mean. The commutativity problem disappears, so that (\ref{eq:wrong}) is now correct, and only the eigenvalues vary over time. This model may or may not be appropriate, depending on the context, as we discuss later.

An obvious choice of dynamics is the CIR process,
\begin{equation}
dX_t = \kappa(\Xbar - X_t ) \, dt + \sigma \sqrt{X_t} \, dW_t.
\label{eq:cir}
\end{equation}
\notthis{The stationary distribution of $X_t$ is Gamma(\footnote{pdf $\Gamma(\nu)\inv \xi^{-\nu-1} x^\nu  e^{-x/\xi}$ where $\nu$ is the shape parameter: the mean is $\nu\xi$ and the variance is $\nu\xi^2$.}) of mean $\Xbar=1$ and shape parameter $2\Xbar\kappa/\sigma^2$.}
Importantly, the joint distribution of $X_t$ and $\exp\big(\int_0^t X_s \, ds \big)$ is known via the double Laplace transform \cite[Prop.~2.5]{Lamberton12}:
\begin{equation}
\ex_0 \left[ e^{-\lambda X_t -\mu \int_0^t X_s \, ds} \right] = \mathcal{A}(\lambda,\mu,t) e^{- \mathcal{B}(\lambda,\mu,t) X_0} 
\label{eq:cirsol}
\end{equation}
with
\begin{eqnarray*}
\mathcal{A}(\lambda,\mu,t) &=&  \left( \frac{2\mykappa_\mu e^{(\mykappa_\mu+\kappa)t/2}}{\sigma^2 \lambda (e^{\mykappa_\mu t}-1)+\mykappa_\mu-\kappa+e^{\mykappa_\mu t}(\mykappa_\mu+\kappa)} \right) ^{2\kappa \Xbar/\sigma^2}
\\
\mathcal{B}(\lambda,\mu,t) &=& \frac{\lambda\big(\mykappa_\mu+\kappa +e^{\mykappa_\mu t}(\mykappa_\mu-\kappa)\big) + 2 \mu(e^{\mykappa_\mu t}-1) }{\sigma^2 \lambda (e^{\mykappa_\mu t}-1)+\mykappa_\mu-\kappa+e^{\mykappa_\mu t}(\mykappa_\mu+\kappa)} 
\end{eqnarray*}
and $\mykappa_\mu=\sqrt{\kappa^2+2\sigma^2\mu}$.
Accordingly the $T$-period transition matrix conditional on $X_0$ is\footnote{The functions $z\mapsto\mathcal{A}(0,z,T)$ and $z\mapsto\mathcal{B}(0,z,T)$ are regular for $\Real z\ge 0$, so can be evaluated `at' $z=-\Gbar$.}
\begin{equation}
\ex_0\left[ e^{\Gbar \int_0^T X_t\, dt } \right] = \mathcal{A}(0,-\Gbar,T) e^{- \mathcal{B}(0,-\Gbar,T) X_0 };
\end{equation}
the unconditional one is obtained by integrating $X_0$ out, which is immediate because its unconditional distribution is Gamma of mean $\Xbar$ and shape parameter $2\kappa\Xbar/\sigma^2$.

\subsection{Historical calibration with a time-varying generator}
\label{sec:tvh}

Default rates fluctuate over time, and we wish to capture this effect in a model that can then be linked to suitable econometric indicators \cite{Fei13}. Thus we need a stochastically-driven model, as opposed to a deterministic but time-inhomogenous one as considered in \cite{Bluhm07}.

As just mentioned, a simple scaling of the generator matrix can be done easily. However, it is unrealistic, because in benign years there are more upgrades and fewer downgrades, and the reverse in bad ones: it is not correct simply to scale both by the same factor. Returning to (\ref{eq:G2}), write
\[
G_\lgen = G_\lgen^+ + G_\lgen^-
\]
in which the terms on the RHS are valid generator matrices that are, respectively, lower- and upper-triangular. These are generators for an upgrade-only and a downgrade-only Markov chain, and so this decomposition is unique. Except in trivial cases we will always have $G^+G^-\ne G^-G^+$.
Now scale these two terms by different positive variables $X^+_t$ and $X^-_t$ representing the contribution from upgrades and downgrades respectively, so that
\begin{equation}
G_{\lgen,t} = X^+_t G_\lgen^+ + X^-_t G_\lgen^-.
\end{equation}
It is easiest to do this in a discrete-time setting, because our natural source of data for calibration is annual. Then the one-year transition matrix is the exponential of $G_{\lgen,t}$, but we no longer have a convenient closed-form calculation for it, and have to resort to (\ref{eq:exp}).
\notthis{Another idea is to scale the subdiagonal elements of $H$ by $X^+_t$ and the subdiagonal elements by $X^-_t$: this generates very similar results.}

We do not have yearly transition matrices to calibrate to: even if we did, they would be quite `noisy', so that for example we might have an instance of a transition A to BB, but none to BB+. However two pieces of annual information are available \cite[Table~6]{SandP19}: the downgrade:upgrade ratio and the default rate, both of which are averaged across rating states. Roughly speaking the former quantity gives information about $X^-_t/X^+_t$, and the latter about $X^-_t$. They can be matched exactly, and uniquely, by a simple numerical search, except in singular cases where there are no upgrades/downgrades/defaults. Doing this for each year gives Figure~\ref{fig:history}.

The sample statistics of $X^+$ and $X^-$ are: mean 0.95, 0.95; standard deviation 0.42, 0.65; correlation 0.55.
This shows, as anticipated, that it is necessary to have two factors rather than just a single one. In fact, this is a similar conclusion to that in \cite[\S2]{Andersson00}. It is perhaps surprising that the correlation is positive, because `good years' should see $X^+$ high and $X^-$ low and oppositely in bad years, which was the whole point of having two factors. The author is indebted to M.~van Beek for pointing out that over shorter time scales, data suggest a negative correlation. In other words, looking at all of 2009 is unhelpful as it is a year of two parts: first multiple downgrades and defaults, then multiple upgrades.
It is hard to be precise about what stochastic process $X^\pm_t$ follow, as we have only 38 data points, but a sensible model choice for each of $X^\pm_t$ is the exponential of an autoregressive (AR) process. That is to say, we write $x_t$ for the logarithm of $X_t$, with the mean subtracted; then the AR($p$) model is 
\[
x_t = -\sum_{j=1}^p a_j x_{t-j} + e_t
\]
where $\{e_t\}$ is a white noise process and the $(a_j)$ are coefficients to be determined. The properties of these processes, and the fitting of their coefficients, are well understood, e.g.\ \cite{Makhoul75,Marple87}.
An AR(1) process captures mean-reversion. An AR(2) process allows more subtlety, and allows the possibility of rating momentum, in the sense that next year's up/downgrade rate is likely to be higher than this year's if this year's was higher than last year's, or in equations
\[
\ex[(x_n - x_{n-1})(x_{n-1}-x_{n-2})] > 0, 
\]
a condition that can be expressed in terms of the autocovariance function and thence of the $(a_j)$ by the Yule-Walker equations. However, there is little evidence for this in the data used here, suggesting that if rating momentum exists it is likely to be observed over a faster time scale than a year, and is possibly more of an idiosyncratic effect, in the sense that troubled issuers often suffer successive downgrades over a period of months (which does occur).

\begin{figure}
\centering
\scalebox{0.75}{\INCLUDE{mig3_coeffs}}
\mycaption{
Evolution of coefficients $X^\pm_t$ in time-varying model.
The green trace $X^+_t$ controls upgrades and the red trace  $X^-_t$ controls downgrades.
}
\label{fig:history}
\end{figure}


\section{Risk-neutral calibration}
\label{sec:rnc}

\subsection{Theoretical considerations} 

Risk-neutral calibration is the identification of a generator matrix that fits a given set of CDS spreads or bond prices, via the survival curve\footnote{The valuation of CDS and bonds in terms of the survival curve is well-understood, e.g.~\cite{OKane08}.}. This is an entirely different problem from fitting to a prescribed historical transition matrix.

The first point we make is that in reality we should not expect anywhere near a perfect fit in doing this: there is no reason why the market should trade a credit in line with its public rating, as the market provides a current view of the firm's future solvency, and the rating may not be up to date. Thus it is quite possible to have two different BBB$-$ firms trading at 80bp and 140bp at the 5y tenor; once we get down to CCC+ we can have names trading anywhere between 500bp and 5000bp. Also there may be no bonds of a particular rating: this will almost certainly be true when rating modifiers are used. All these examples can be found in the dataset we used. Therefore we must be prepared for very `noisy' data, and the objective of fitting must be to come up with sensibly-shaped curves that give a reasonable fit.

It is superficially attractive to describe this as a `bond pricing' model. However, bonds are quoted on price or spread-to-Treasury, so there is no need for a pricing model as such. Arguably it is suitable for `matrix pricing' of illiquid bonds, but an illiquid bond is likely to trade at a significant yield premium to a liquid one of the same rating and tenor, so this would have to be borne in mind. In reality, the model is most suited to relative value analysis or for a parsimonious representation of a certain sector, e.g.~``Where do US industrials trade for different rating and maturity?'' Another application is as an input to the evaluation of EM corporate bonds, coupling the developed market spread to the country spread \cite{Martin18c}.

Our next point is different, but no less fundamental. Even in a hypothetical world of bond/CDS spreads exactly lining up with their public ratings, and trading with great liquidity and low bid-offer, does the term structure, for each rating, allow us to uniquely identify the generator? There is a clear intuition about why the term structure on its own does not suffice.  Knowing the full transition matrix allows us to value any contingent claim, including claims that require us to know about rating (and hence price) \emph{volatility}. But such information cannot be gleaned from bond prices alone.

The term structure provides (in principle) $2n$ clearly-interpretable pieces of information, in the form of the short-term spread and the gradient of the spread curve at the short end, but in practice observations of these quantities will be `noisy'.
Denoting by $Q_i(T)$ the survival probability for time $T$, conditional on being in state $i$ at time zero, we have for small $T$,
\[
Q_i(T) = 1 - G_{i\maltese} T - (G^2)_{i\maltese} \frac{T^2}{2}  + O(T^3).
\]
If we convert to a par CDS spread then the PV of the payout leg and the coupon leg are, for small maturity $T$, respectively
\[
(1-\recov) \big(1-Q_i(T)\big) = (1-\recov) \left( G_{i\maltese} T + (G^2)_{i\maltese} \frac{T^2}{2} + \cdots \right) \]
(with $\recov$ the recovery rate), and
\[
 T \frac{1+Q_i(T)}{2} = T \left( 1 - G_{i\maltese} \frac{T}{2} + \cdots \right) 
\]
and the CDS spread is the ratio of the two:
\[
\frac{s_i(T)}{1-\recov} = G_{i\maltese} +  \big[ (G^2)_{i\maltese} + (G_{i\maltese} )^2 \big]  \frac{T}{2} + \cdots
\]
Alternatively we can use the continuously-compounded zero-coupon Z-spread  $s_i(T) = -T\inv \ln Q_i(T)$, and the result is the same.
Writing the matrix square $G^2$ as an explicit sum, we obtain after some elementary algebra:
\begin{equation}
\begin{array}{rcl}
s_i(0) &=& (1-\recov) G_{i\maltese} \\
s_i'(0) &=& \displaystyle \frac{1-\recov}{2} \sum_{j=1}^n G_{ij} (G_{j\maltese} - G_{i\maltese})
\end{array}
\label{eq:y0}
\end{equation}
The first of these is obvious: the short-term spread is explained by the instantaneous probability of transition to default. The second expresses extra loss arising from transition to riskier states ($i\to j$ with $G_{j\maltese}>G_{i\maltese}$) balanced off against transition to less-risky states ($G_{j\maltese}<G_{i\maltese}$): if the former exceeds the latter, the spread curve is upward-sloping, and vice versa.
In particular, the highest rating necessarily has an upward-sloping curve and the lowest a downward-sloping one.

Now suppose we have two term structures that agree on $s_i(0)$ and $s_i'(0)$ for each $i$. They might look different at the long end ($T\to\infty$), but how many extra parameters would such a difference entail? So it seems plausible that the term structure provides a little over $2n$ pieces of information, and certainly nowhere near $n^2$.
A `toy' example shows this well (Figure~\ref{fig:tm3}\footnote{The letter codes A,B,C do not correspond to S\&P ratings: they are just labels of convenience, with A the best and C the worst. Recovery $\recov=0.4$.}). The term structures are not identical, but are so close that there is no realistic way of telling the generators apart---we chose, as is easily done, matrices for which the information in (\ref{eq:y0}) is identical. Rating B clearly has higher volatility in the second matrix.

\subsection{Spectral theory}
\label{sec:spectral}

It is possible to use basic differential geometry to provide some quantitative precision about invertibility. There exists a smooth \emph{pricing map} $f\in\mathcal{C}^\infty$ from the space of generator matrices $\mathcal{G}$ to the space of term structures $\mathcal{S}$, that is, a set of curves of spread vs maturity, one for each rating. 
At any point $G\in\mathcal{G}$ it has a derivative $\partial f$ that relates, linearly, infinitesimal changes in generator matrix to infinitesimal changes in term structure, and represented by the jacobian\footnote{The jacobian of a map $f:\mathcal{X}\to\mathcal{Y}$ at $x\in\mathcal{X}$ is the matrix $J_f(x)$ whose $(i,j)$th element is $\partial f_i/\partial x_j$ evaluated at $x$.} $J_f$. To construct it, take $G\in\mathcal{G}$ and perform, one by one, $n^2$ bumps on it, in which each bump consists in increasing the element $G_{ij}$, for $i\ne j$, by a small amount $\epsilon$ while adding $-\epsilon$ to $G_{ii}$ so as to keep the $i$th row sum zero. Compute the new term structure $f(G+\delta G)$, written as a set of $Nn$ points (maturities 1y, 2y, \ldots $N\,$y and $n$ ratings). The difference $\big(f(G+\delta G)-f(G)\big)/\epsilon$ estimates the directional derivative, and repeating this for each possible bump gives the jacobian, which is of dimension $Nn \times n^2$, containing the sensitivity of all points on the term structure to all $n^2$ bumps.
The degree to which $f$ is locally invertible\footnote{The singular values of $J_f$ only tell us about \emph{local} invertibility: it is easy to construct an $f$ that causes the domain to be folded on itself in such a way that $J_f$ is nonsingular but $f$ is not one-to-one. A simple example is $f:\R^2\to\R^2$ given by $f(x,y)=e^x(\cos y, \sin y)$: the singular values of $J_f$ are identical, being $e^x$, but $f$ is periodic in $y$, and so obviously not one-to-one. In practice, therefore, establishing that a function $f:\R^d\to \R^d$ is invertible is hard if $d>1$. In context this means that we might have two totally different generator matrices giving precisely the same term structure, though this is perhaps unlikely.} can be understood by examining the singular value decomposition (SVD, \cite[\S2.6]{NRC}) of the jacobian, that is, $J_f= U\Delta V'$, in which $U,V$ are orthogonal and $\Delta=\mathrm{diag}(\delta_1,\delta_2,\ldots,)$ with $\delta_1\ge \delta_2\ge\cdots\ge0$: these elements are the roots of the characteristic equation $\det(\delta^2 I-{J_f}'J_f)=0$, and are the \emph{singular values}. In attempting to locally invert $f$ we have to divide by $\Delta$, so small singular values cause a problem. As an overall scaling of the matrix does not affect its conditioning, standard procedure is to plot $\hat{\delta}_i=\delta_i/\delta_1$, for $i=1,2,\ldots$, on a logarithmic scale, and we call this the \emph{normalised spectrum} of $\partial f$.
In well-conditioned problems the $\hat{\delta}_i$ are all roughly equal; in ill-conditioned problems some are very small, and in these coordinate directions inversion will be impossible without grossly amplifying any observation noise. 

Taking one further step we also want to examine whether a particular parametrised subspace of $\mathcal{G}$ permits stable inversion: in context $\mathcal{P}$ is, of course, the TDST parameters.
Let $\mathcal{P}$ be the space of parameters, with a map $\pi:\mathcal{P}\to\mathcal{G}$ taking a particular parameter set to its associated generator. We can construct $J_\pi$ and $J_{f\circ\pi}$ by bumping the parameters $p$ and seeing the effect on the generator $\pi(p)\in\mathcal{G}$ and the term structure $f(\pi(p))\in\mathcal{S}$. Notionally the singular values of the restricted\footnote{i.e.\ the restiction of $\partial f$ to the tangent space of $\pi(\mathcal{P})\subset \mathcal{G}$.}  $\partial f$ are simply those of $J_{f\circ\pi} \cdot J_\pi\inv$, but this is not meaningful as $J_\pi$ is a rectangular matrix: instead, the values we seek are the roots $(\delta_i)$ of the more general characteristic equation
\begin{equation}
\det \big( \delta^2 {J_\pi}' J_\pi  - {J_{f\circ\pi}}'J_{f\circ\pi} \big) =0 .
\label{eq:secular2}
\end{equation}
This construction is invariant under local reparametrisation: if we have two different parametrisations of the same subspace of $\mathcal{G}$, i.e.\ $\pi_1:\mathcal{P}_1\to \mathcal{G}$ and $\pi_2:\mathcal{P}_2\to \mathcal{G}$, with $\pi_1\inv\circ \pi_2$ locally invertible, then the above equation does not depend on the choice $(\pi_1,\mathcal{P}_1)$ or $(\pi_2,\mathcal{P}_2)$: in other words it just depends on $f$ and on the subspace of $\mathcal{G}$.

\subsection{Empirical work} 

\begin{figure}
\centering
\scalebox{.75}{\includegraphics*{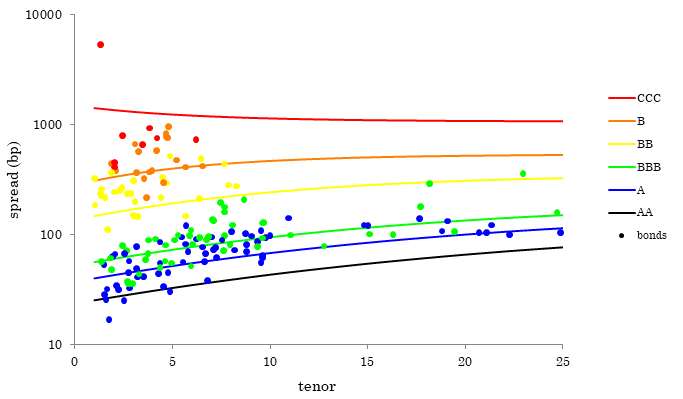}} 
\mycaption{Implied par spreads from fitting to US bond data (manufacturing sector, 05-Feb-20).
}
\label{fig:spds}
\end{figure}

We fitted an 18-state matrix to US bond data from the manufacturing sector on 05-Feb-20. To reduce the dimensionality of the problem further we used the same trick as in \S\ref{sec:hc}, specifying sub- and super-diagonal elements for AA,A,BBB,BB,B,CCC and dealing with rating modifiers by interpolation: this gives 14 parameters ($=2\times 6 + n_\lgen$).
The performance surface was quite flat: different parameter sets gave almost the same quality of fit. We have more to discuss, but the reader may wish to glance at Figure~\ref{fig:spds}, which shows the end result. To avoid cluttering the plot, rating modifiers are not shown, so that for example BBB+,BBB,BBB$-$ are all coloured the same as BBB.
One obvious feature is the high degree of dispersion of bond spreads around their fitted curves.
We can also see from Figure~\ref{fig:spectra}(a) that there is no realistic way of fitting anywhere near $n^2=324$ parameters, as the singular values roll off so rapidly. This first graph can be constructed for any generator so it has nothing to do with TDST.
Turning next to Figure~\ref{fig:spectra}(b), which pertains to TDST, we see from the green trace that some instability is likely as the smallest normalised singular values are quite small. This points to the conclusion that further constraining is required. We have already anticipated that volatility assumptions need to be incorporated, and so we address that next.

\subsection{Adding spread volatility}

If we wish to capture spread volatility in a realistic way, we should augment the model to allow for the generator to be stochastic, thereby capturing day-to-day fluctuations in spread that are not associated with rating transition. Let us therefore return to \S\ref{sec:tvtheor} and revisit the idea of a CIR process to drive this. Equation (\ref{eq:cirsol}) giving the joint law of $X_t$ and $\exp(\Gbar \int_0^t X_s \, ds)$ allows a wide range of contingent claims to be valued. However, single-name CDS options scarcely trade\footnote{See \cite{Martin11c} for a review and discussion of this subject.} so there would be nothing to calibrate to.

A more practical idea is to compute the instantaneous spread volatility and attempt to match it to historical data. As the effect of $X$ is approximately to scale the credit spreads proportionally, the instantaneous volatility of the period-$T$ spread for rating $i$ is
\begin{equation}
\sigma^s_i(T) \approx \left( \sum_{j=1}^n G_{ij} \big( \ln s_j(T) - \ln s_i(T) \big)^2 + \sigma^2 \right)^{1/2}
\end{equation}
where $\sigma$ is the volatility parameter in the CIR process.

Turning now to empirical matters, we make some simple observations, based on time series analysis of some 1500 bond and CDS spreads going back 10--20 years. The first is that credit spread volatility is roughly 40\%, and that this does not depend strongly on rating, tenor, or spread level. The second is that there is little evidence of mean reversion, and so $\kappa$ in (\ref{eq:cir}) needs to be quite low: we have fixed it at 0.1(/year), though the model calibration is not sensitive to this choice. Thus, at the expense of adding one more\footnote{As $\Xbar$ can be set to unity.} parameter ($\sigma$), we can penalise the deviation of each $\sigma^s_i(T)$ from our assumed value of 0.4. This was found to stabilise the calibration. 

We can now redo the spectral analysis of Figure~\ref{fig:spectra}, augmenting the pricing map $f$ so that it takes a given generator $G\in\mathcal{G}$ to the term structure of spread and also of instantaneous volatility, notated $f:\mathcal{G}\to\mathcal{S}\times\mathcal{V}$. The singular values, as expected, roll off more slowly (blue trace), showing that new information is being supplied by the volatility term structure. By Figure~\ref{fig:spectra}(a) this is, unsurprisingly, still not enough to allow 324 parameters to be fitted, but Figure~\ref{fig:spectra}(b) suggests that enough stability may be conferred within the TDST parametrisation, which is what we found in practice.

A final point about this kind of model is that it is capable of capturing the effects seen in early 2008 where high-grade credits had inverted yield curves even at modest spreads of 100--200bp. As we said earlier, a static generator cannot deal with this as the spread curve for the best rating state cannot possibly be inverted, and in any realistic parametrisation all the investment-grade curves will be not be. But with a dynamic generator this is now possible, if one introduces $X_0$ into the calibration. When it is very much higher than its long-term mean, short-term spreads will be elevated, as happened in that time period.

\begin{figure}

\begin{tabular}{ll}

\begin{tabular}[t]{l|rrrr}
 & A & B & C & $\maltese$ \\
\hline
A& $-0.14$ & 0.10 & 0.03 & 0.01 \\  
B& 0.00 & $-0.05$ & 0.00 & 0.05 \\  
C& 0.00 & 0.05 & $-0.15$ & 0.10 \\  
\end{tabular}

& 

\raisebox{-0.8\height}{\scalebox{0.625}{\INCLUDE{mig3_ts3a}}}

\\

\begin{tabular}[t]{l|rrrr}
 & A & B & C & $\maltese$ \\
\hline
A& $-0.14$ & 0.10 & 0.03 & 0.01 \\  
B& 0.20 & $-0.41$ & 0.16 & 0.05 \\  
C& 0.00 & 0.05 & $-0.15$ & 0.10 \\  
\end{tabular}

&

\raisebox{-0.8\height}{\scalebox{0.625}{\INCLUDE{mig3_ts3b}}}

\end{tabular}

\mycaption{Two different hypothetical generators (last row omitted) and corresponding term structures, which are seen to be virtually identical. }

\label{fig:tm3}
\end{figure}

\begin{figure}
\centering
\begin{tabular}{ll}
(a) \scalebox{0.625}{\INCLUDE{mig3_tm18_jac}} &
(b) \scalebox{0.625}{\INCLUDE{mig3_tm18_par_jac}} 
\end{tabular}

\mycaption{Normalised spectrum $\hat{\delta}_i$ vs $i$ of $\partial f$ (the derivative of the pricing map), using the model calibration for Figure~\ref{fig:spds}.
In each case ``spreads'' means $f:\mathcal{G}\to \mathcal{S}$, and ``spreads+vols'' means $f:\mathcal{G}\to \mathcal{S}\times\mathcal{V}$.
Plot (a) refers to the map $f$, and (b) to its restriction to $\pi(\mathcal{P}) \subset \mathcal{G}$, using (\ref{eq:secular2}) to compute the singular values.
}

\label{fig:spectra}
\end{figure}


\newpage

\section{Conclusions}

We have presented a new model (TDST) based on a tridiagonal generator matrix coupled with a stochastic time change to allow multiple downgrades to occur with a probability that is not far too low. The model may be calibrated historically or to bond/CDS spreads, and the relevant conclusions are given below. An incidental advantage is that the matrix exponential is generally easy to calculate.

With regard to historical calibration, the presence of a `WR/NR' (unrated or withdrawn rating), and the fact that some transitions are very rare, gives rise to considerable uncertainty as to the true underlying transition probabilities. The TDST model as described here is sufficiently flexible to calibrate within these uncertainties, without having too many parameters. If the model is made time-varying, two factors are needed to capture the volatility of upgrades and downgrades.

With regard to risk-neutral calibration, it is impossible to unambiguously identify a generator matrix from term structure, and two ingredients are essential: (i) dimensionality reduction e.g.\ through TDST and (ii) information about volatility. Once this is provided the calibration works well even in the presence of real data which must be expected to be very `noisy'.

Using the theory of \S\ref{sec:spectral} to analyse the derivative of the pricing map should be a standard procedure in any calibration exercise. It is not at all restricted to the application in this paper.


\section*{Acknowledgements}

I thank Di Wang and Huong Vu for their work in earlier stages of this project, and Roland Ordov\`as and Misha van Beek for helpful discussions.

\appendix
\section{Appendix}
\subsection{Comment on the Gamma process}

In the limit of one or more super- or sub-diagonal elements of $H$ being zero, i.e.\ an intransitive generator, the $H=D\widetilde{H}D\inv$ construction will no longer work, because $D$ becomes singular. However, in the case of the Gamma model the transition matrix is simply a matrix power, as it is
\[
\exp(\lgen(H)t) = (I-H/\beta)^{-\beta t}.
\]
Going back to (\ref{eq:exp}), we can now see why the second expression is a valid transition matrix.
It can be computed directly, as follows. First, compute $A=(I-H/\beta)\inv$, which is most easily done by LU factorisation of $I-H/\beta$. Next, raise this to the power $\nu=\beta t$. As raising a matrix to an integer $\nu$ is easy (by successive squaring and using the binary representation of $\nu$), we only have to deal with the case where $\nu\in(0,1)$.
Applying the binomial theorem twice and interchanging the order of summation:
\begin{eqnarray*}
A^\nu &=& (I+(A-I))^\nu \\
&=& \sum_{r=0}^{m} \frac{\Gamma(\nu+1) (A-I)^r}{\Gamma(\nu+1-r)\,r!} + \epsilon_m \\
&=& \sum_{r=0}^{m} \frac{\Gamma(\nu+1)}{\Gamma(\nu+1-r)\,r!} \sum_{k=0}^r (-)^{r-k} {r \choose k} A^k  + \epsilon_m \\
&=& \sum_{k=0}^{m} A^k c_k(\nu) + \epsilon_m  \\
c_k(\nu) &=& \frac{1}{k!} \sum_{r=k}^{m} \frac{(-)^{r-k} \Gamma(\nu+1)}{(r-k)! \, \Gamma(\nu+1-r)} \\
\end{eqnarray*}
where $\epsilon_m$ denotes the truncation error.
It is easy to see that $c_k(\nu)$ is a polynomial of degree $m$ that vanishes at $\nu=0,1,\ldots,k-1,k+1,\ldots,m$, while $c_k(k)=1$. (See \cite[\S0.154]{Gradshteyn94}.)
Therefore this method writes $A^\nu$ as the Lagrange interpolant between $I,A^1,\ldots,A^m$; the simplest formula is $m=1$ which gives a linear interpolation between $I$ and $A$. The expansion is convergent because in context the eigenvalues of $A$ lie in $(0,1]$.

\bibliographystyle{plain}
\bibliography{}

\end{document}